Dr.-Ing. Janine Glänzel[1]
Andreas Naumann[2]
Tharun Suresh Kumar [1*]

# PARALLEL COMPUTING IN AUTOMATION OF DECOUPLED FLUID-THERMOSTRUCTURAL SIMULATION APPROACH

Decoupling approach presents a novel solution/alternative to the highly time-consuming fluid- thermal-structural simulation procedures when thermal effects and resultant displacements on machine tools are analyzed. Using high dimensional Characteristic Diagrams (CDs) along with a Clustering Algorithm that immensely reduces the data needed for training, a limited number of CFD simulations can suffice in effectively decoupling fluid and thermal-structural simulations. This approach becomes highly significant when complex geometries or dynamic components are considered. However, there is still scope for improvement in the reduction of time needed to train CDs. Parallel computation can be effectively utilized in decoupling approach in simultaneous execution of (i) CFD simulations and data export, and (ii) Clustering technique involving Genetic Algorithm and Radial Basis Function interpolation, which clusters and optimizes the training data for CDs. Parallelization reduces the entire computation duration from several days to a few hours and thereby, improving the efficiency and ease-of-use of decoupling simulation approach.

## 1. INTRODUCTION

When a machine tool is subjected to changes in environmental influences such as ambient air temperature, velocity or direction, then flow (CFD) simulations are necessary to effectively quantify the thermal behaviour between the machine tool surface and the surrounding air (fluid), ref Glänzel et al. [1]. This two-step simulation procedure involving fluid and thermo-structural simulations is highly complex and time-consuming especially when complex geometries or dynamic components are considered. Reducing the dependency of thermo-structural simulations on CFD simulations for convection data is a matter of great concern as far as reduction in computation time is concerned. A suitable alternative for the above process can be attained by decoupling CFD and thermo-structural simulations .This can be achieved by introducing a clustering algorithm (CA) and characteristic diagrams (CDs) in the workflow.

---

[1] Fraunhofer Institute for Machine Tools and Forming Technology IWU Chemnitz, Germany

[2] Dresden University of Technology, Institute of Scientific Computing, Dresden, Germany

* E-Mail: Tharun.Suresh.Kumar@iwu.fraunhofer.de

 DOI: xxxxxxxx ,

CDs are continuous maps of a set of input variables onto a single output variable and consists of multidimensional grids, refer to the work by Priber [2] and Naumann et al. [3]. Using a limited number of CFD simulations, the environmental influences at certain optimal node points on the machine surface can be mapped on to its corresponding heat transfer coefficient (HTC) values which serve as the training data for CDs. Later on, these CDs can be used to interpolate convection data as and when required for a user-defined input load case. This eliminates the need to perform CFD simulations each time an environmental boundary condition is varied. However, there was still scope for improvement in the time required to train the CDs. Training involves two steps:

• Running a certain number of CFD simulations and exporting the HTC values on the machine tool surfaces at each load case (combinations of environmental influences). HTC is an important parameter, which effectively defines the amount of heat transfer per unit area on a solid body for a specific temperature difference between the solid face and the surrounding fluid area.

• Finding optimal subsets of FE-node points using a clustering technique which basically involves Genetic Algorithm (GA) and Radial Basis Function (RBF) interpolation. Optimal subsets of FE-nodes are found on each face of the machine using GA such that, HTC values when interpolated using RBFs over a machine face using these optimal node points will have the least possible error. A particular optimal subset is evolved gradually for a face over each iteration/ generation in GA. Each CD corresponds to a single optimal node point where the parameterized environmental influences are mapped directly onto corresponding HTC values.

Both these operations are significantly time consuming depending on the accuracy of results required. A single CFD simulation could take around one hour to complete. The realistic discretization or proper training of CDs would involve 158 CFD Simulations, ref (Glänzel et al. [1]). Thus, approximately six days would be required to perform CFD simulations on a normal computational system. The rate at which optimal node points on machine tool surfaces are found depends on the complexity of the faces and the GA parameters chosen. Approximately, for a surface of 2000 FE-node points around 10000 generations are required to effectively optimize the FE-node subset. This process also involves several hours of computation. Both these operations could in combination require ten to fifteen days of computation for a machine tool. A possible remedy to this problem is proposedly parallelization.

Nowadays, parallel processors and systems are extensively used. Even the consumer CPUs consist of at least two cores, or are able to execute at least two threads concurrently. The efficiency of parallel computations depends highly on the data dependency between different tasks and the distribution of the workload for the tasks. A complicated (and dense) data dependency leads to communication (overhead) and should be avoided for an efficient work partitioning strategy. This problem is circumvented by structuring the work parameters and deducing independent tasks.

If the runtime of every task varies drastically, i.e. by magnitudes, the work has to be balanced among the available processors. For very different workloads, this strategy leads to a very inefficient CPU utilization and a load balancing strategy has to be employed. Even though the CFD simulations are expensive, a very balanced runtime was observed between the parameters. In contrast, the clustering algorithm leads to very different work loads, such that a heuristic number partitioning algorithm is employed to balance the work.

The structure of parallelization adopted in decoupling approach is explained in the next chapter. Chapter 3 deals with the first type of parallelization i.e. parallel computation of CFD simulations and thermal data extraction. The second type of parallelization is explained in chapter 4. It involves parallel computation of clustering algorithm which is used to reduce the training data for Characteristic Diagrams. In chapter 5, a case study on a valid FEM model is performed which illustrates the reduction in computation time using both parallelization strategies. A brief summary and future scope of work of this approach are discussed in chapter 6.

## 2. APPROACH - DECOUPLING AND PARALLELIZATION

The decoupling approach (refer to the work by Glänzel et al [4]) eliminates the dependency of thermo-mechanical simulations on CFD simulations by introducing Clustering Algorithm (CA) and CDs in between the two simulation workflows as shown in figure 1, refer Glänzel et al. [5]. The HTC values are exported as comma separated value (*.csv) files for each machine face under varying ambient load cases (air temperature, air speed and directions of air flow) from ANSYS-CFX. Each simulation corresponds to a particular combination of environmental influences or a load case. The first stage of parallelization, as shown in figure 2 can be incorporated in this scheme of activities, which involves simulations and data export. All or most of the simulations can be run in parallel, drastically reducing the cumulative time needed for CFD simulations. This facilitates exponentially fast transfer of HTC data for clustering and thereby training the CDs.

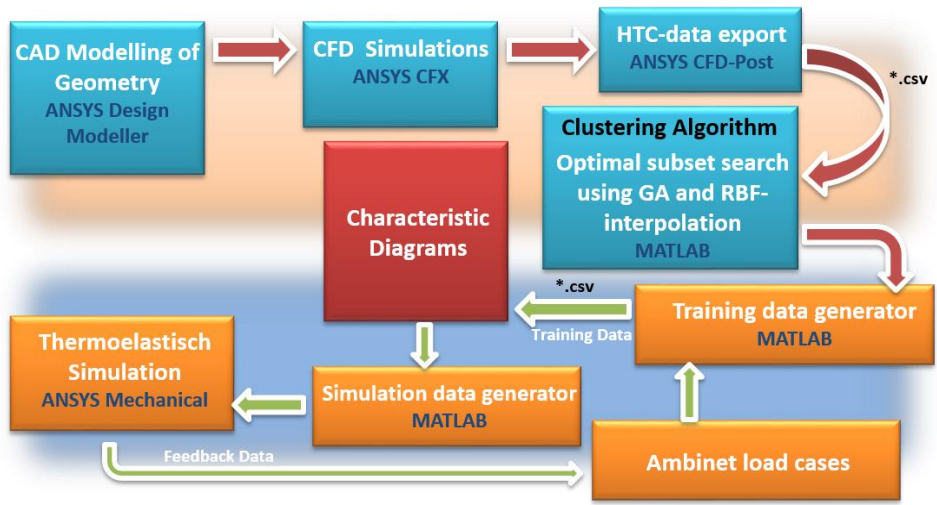

Fig. 1. Decoupling workflow

The basic function of CA is to reduce the vast amount of data needed to train the CDs and thereby reduce the computation time. CA performs optimal subset search of FE-nodes on the machine tool surfaces using an optimization algorithm (GA) and RBF interpolation. Both operations are performed using MATLAB scripts. GA involved finds the optimal subset of nodes on each/selected machine tool surface(s) based on an objective or fitness function. The fitness of a particular subset corresponds to the interpolation error between the actual HTC values (exported from CFX ) and RBF- interpolated (using this subset) HTC values. Towards the end of a certain number of generations in GA, the subset of FE-node point with the smallest (best) fitness value will be the optimal subset. The size of the subset can be defined by the user based on the accuracy and computation time expected.

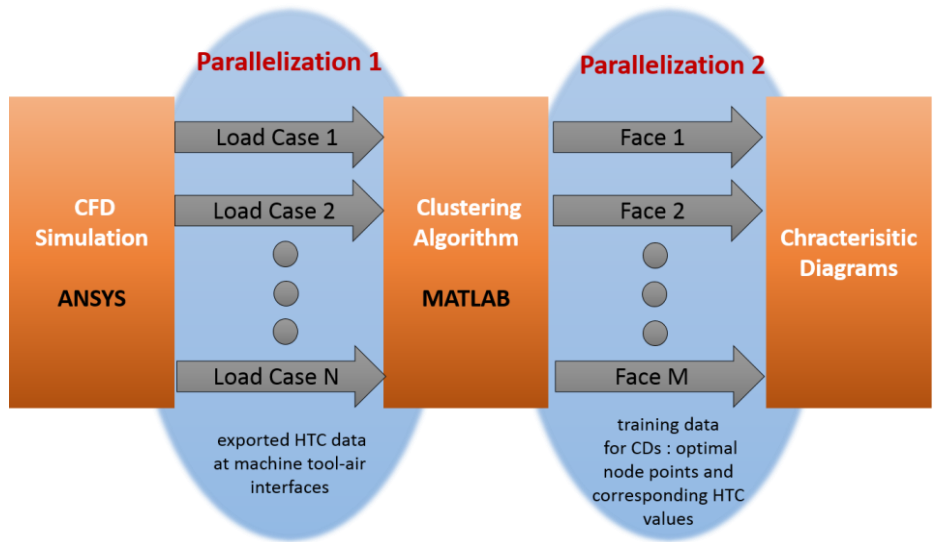

Fig. 2. Parallelization in Decoupling

The clustering operation paves way to the second stage of parallelization as shown in figure 2. The optimal subsets are found on each machine tool surface (air-machine interface). Based on the complexity of the machine surface, the search space for GA could have (on an average) about 1000 FE-node points. The optimization operation here involves two steps for each face. Optimal subsets are found for each load case in the first step. From these subsets, a final optimized subset is deduced which will be utilized for training CDs irrespective of the air temperature, velocity and flow directions. Thus, clustering can be a significantly time-consuming if the search space is too large. Parallelization 2 as shown in figure 2 can be utilized here and optimal subsets can be simultaneously obtained at each machine face. Later on, CDs are formulated for each optimal node points which are capable of predicting HTC values for user-defined load cases based on the training data. The predicted HTC values for the optimal subsets serve are RBF points used to interpolate the HTC for the entire faces of the machine. This serves as the boundary (convection) data for thermo-mechanical simulations.

## 3. PARALLELIZATION IN CFD SIMULATIONS

One of the most famous laws in computational sciences is Moores law [6], which states that roughly every 18 months the computing capability doubles. In the last twenty years this observation switched from single-core capabilities to multi core systems. Nowadays, all major CPUs contain at least two cores. Multi-core systems, with more than 12 cores, are also available. From this point of view recent software packages have to support multi core systems to retain their efficiency. Furthermore the near future also heads to Exascale computing, refer to the investigation by Markidis et al [7].

Speaking of parallel architectures, in the sense of software as well as hardware, there are two main categories. The so called shared memory systems belong to the first category. The attribute shared refers to the memory of the processes running on a single system. Hence every process might (but not have to) access the memory of another process. This makes communication and data exchange efficient. For very tightly coupled algorithms this approach is therefore advantageous. But it comes at the drawback that every process also shares the connections between memory and CPU with the neighboring process. Hence algorithms, which rely on a lot of memory per instruction are often bandwidth bound.

The other category involves the distributed systems. In this case every process runs on its own system and has only direct access to the own memory and CPU. This architecture makes the communication less efficient, but it does not suffer from the memory bandwidth. Hence this approach is especially interesting for parallel approaches with only small couplings and independent processes, refer to the literature survey by Tanenbaum et al [8].

The efficient parallel computation of the HTC for different load configurations requires each simulation to be defined in a unique way. The design points (in ANSYS) are used to identify each experiment and make them independent to each other. This structure allows the simulations to be run concurrently. Furthermore, the access to the (heterogeneous) HPC system with around 47000 processors allows each simulation to be run on distinct sets of processors.

It is widely known that the runtime of CFD computations depend drastically on the numerical and physical parameters. Incidentally, in the experiment under consideration, variable runtimes were not seen. Very similar runtimes were obtained for every design point. Hence, approximately the same resources could be reserved for every simulation without any drawback.

Parallelization 1, as explained in section 2 involves parallel computation of CFD simulations in ANSYS-CFX. A sufficiently complex CAD-geometry of a machine tool within air medium needs to be chosen to effectively interpret the parallelization of decoupling approach. The CAD model should be justifiable, considering the application of parallelization in a moving machine component under varying environmental influences.

The major objective of CFD simulations in CFX is to provide simulation data (basically HTCs with corresponding FE-coordinates) for each face under each combination of environmental parameters such as ambient temperature, air velocity and air flow directions (azimuth and elevation angles). HTCs are static and relatively smooth on any given flat surface. They have a tendency to fluctuate strongly at edges between two surfaces. Therefore, as long as each machine tool face (2D surfaces) is considered independently, the CDs are expected to store and interpolate HTCs with sufficient accuracy, refer Naumann et al. [3]. Depending on the complexity and size of the FE-mesh used, the simulation duration can vary from a few minutes to several hours.

All major FE tools, including ANSYS-CFX, provide at least an "internal" parallelization of the FE discretization. The parallelization in ANSYS CFX utilizes the Single Program Multiple Data (SPMD) concept. The model and mesh is decomposed into several partitions that are executed as independent tasks that periodically exchange data for solution update. Each process executes the same computations as in a serial run mode. In simple words, the work is evenly distributed and load balanced. This widely used approach is extended for the parallel computation of the HTCs for the CDs. Each simulation with a fixed tuple (T$_{air}$, v, az, el), referring to the velocity, ambient temperature and flow directions - azimuth and elevation respectively, computes the HTCs for this setting on all faces. In other words, the simulations are independent for different combinations of environmental influences/parameters.

The independence among load cases facilitates every simulation to be run in parallel. The software package ANSYS-CFX provides two features to accomplish the parallelization. First, it allows the definition of design points. These points represent the physical parameters velocity and ambient temperature. Second, it features an external python interface through ANSYS/ACT. With this interface, the design point selection, simulation and extraction of convection data on all desired faces afterwards, are automated without user interaction.

Both the above mentioned features are combined for parallelization. At first, an ANSYS-CFX python script is prepared, which runs the simulation for the desired design point and stores the HTC values for all desired faces. Afterwards, a shell script is prepared for the batch system, which is parameterized with the design point. This script then runs ANSYS-CFX in batch mode with the python script and the design point ID.

## 4. PARALLELIZATION IN CLUSTERING FE-NODES

4.1 Introduction to Clustering Algorithm

Clustering is the task of grouping a set of objects in such a way that objects in the same group, called a cluster, are more similar (based on the objective) to each other than to those in other groups. Formally, given a data set of '$m$' dimensions and '$n$' points, $D \in R^{\{n,m\}} = \{d_1, \ldots, d_n\}$, clustering is the process of dividing the points up into '$k$' groups (clusters) based on a similarity measure. The search for a universal or more generic search algorithm led to the discussion on GAs. The GA attempts to find a very good (or best) solution to the problem by genetically breeding the population of individuals over a series of generations and effectively overcome local minima based on Darwinian principle of reproduction and survival of the fittest, analogous of naturally occurring genetic operations such as crossover and mutation (refer to the works by John R Koza [9] and Koenig [10]).

Genetic Algorithm was introduced as a clustering technique and implemented in decoupling approach in work by Glänzel et al. [5]. As discussed in the section 2, the purpose of clustering in decoupling approach is to reduce the number of nodes and corresponding HTC values used for training CDs. Maintaining accuracy in interpolation even after reduction of nodes is very important. This is done by choosing optimal subsets of nodes with a fixed size '$m$' of node number values over each face of the machine, which will be used to build an interpolation function, based on RBFs. The GA addresses the 'Optimal Subset Problem' (refer to the work by Unger et al. [11]) by minimizing the weighting function '$f$' as

$$\min_{\substack{S \subset V \\ |S|=m}} f(S) \quad (1)$$

where $V$ is the set of node numbers on a particular face, $V = \{1, 2 \ldots N\}$ which corresponds with nodes $x_1, x_2, \ldots x_N$ of the finite element mesh and the simulated HTC values $w_1, w_2, \ldots w_N$ in these nodes. In the decoupling approach, the weighting function will calculate the interpolation error which occurs when the '$m$' nodes of '$S$' are used to interpolate the HTC values over an entire machine face, such that error measure in pointwise computed form (2), $f(S)$ becomes zero if $m = N$, and becomes greater than zero if $m < N$.

$$f(S) := \max_{i=1\ldots N} |f_s(x_i) - w_i| \quad (2)$$

The major advantage of GA is that it can be used in those situations where the numerical or mathematical models fail. GA, being an evolutionary algorithm, the progress can be viewed with each iteration.

The clustering algorithm is implemented between the CFD simulations and training data generator (script written in MATLAB) for CDs as shown in figure 1. HTC values obtained over all the faces of the machine/ specimen for different load cases (ambient temperature, flow velocity, azimuth and elevation angles) serve as the input data for clustering algorithm. The parameters of GA such as population size, number of genes, crossover and mutation probability and number of generations are specified as user input.

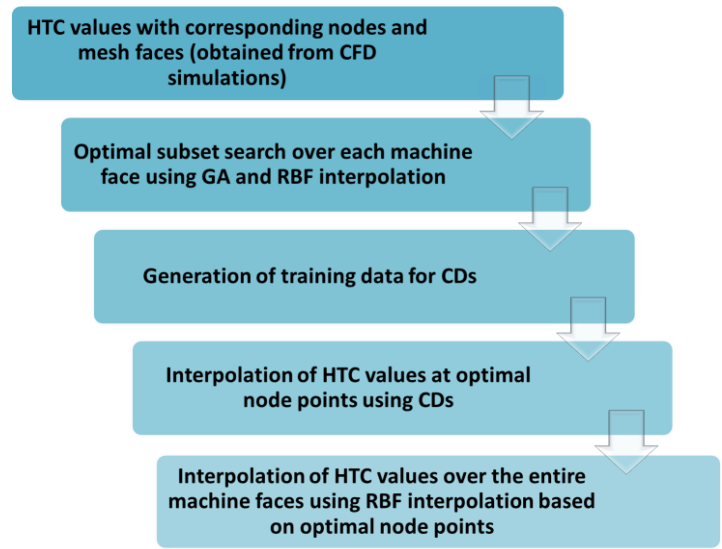

Fig. 3. Implementation of clustering algorithm

CDs are one of the most adopted tools by engineers to approximate real valued functions that depend on one or more input variables. The CDs used in this paper are based on smoothed grid regression technique suggested in the work by Priber [2]. It was later improved to high dimensional CDs which were able to approximate thermo-elastic deformations in machine tools. CDs are continuous maps of a set of input variables onto a single output variable. They consist of a grid of support points along with kernel functions which describe the interpolation in between, refer to the work by Putz et al. [12].

In decoupling approach, training data for CDs are developed using optimal node points. Thus, each CD corresponds to a particular optimal node. HTC values are interpolated over the optimal nodes based on the user defined load cases. Finally, HTC values are interpolated again over the entire faces of the machine using HTC values on the optimal nodes. Thus, the algorithm involves three interpolation processes at different stages of the workflow. RBF interpolation is utilized initially to find the optimal subset and finally to interpolate HTC over all the faces. CDs are used to interpolate the HTCs over optimal node points.

Parallelization 2 basically attempts in running the GA operations simultaneously and thereby finding the optimal node points on each prominent machine tool –air interfaces, which are expected to contribute the most to the thermo-structural displacements on the machine tool and thereby, TCP.

## 4.2 Static Load balancing for the parallelization of the GA operations

Parallelization 2 yields optimal FE-node points at selected machine faces. The training data for CDs is formulated based on these optimal node points such that at each optimal node point, environmental influences such as air flow temperature, velocity and directions of flow (azimuth and elevation angles) are parameterized and mapped onto the corresponding HTC values ($\alpha$). Thus, it would eliminate the need to consider position co-ordinates as input variables in CDs. In the current implementation, air temperature, velocity, azimuth and elevations angles of air flow corresponding to each optimal node point would serve as the input parameters, refer [3]. Therefore, the main aim is to try and quantify the following correlation:

$$(T_{air}, v, az, el) \rightarrow \alpha \qquad (3)$$

The basis for the parallelization 2 (shown in figure 2) is the independence of clustering operation among the machine tool faces. Every face has different optimal node selection, and every selection is independent from the other faces. The evaluation costs of the fitness function depend quadratically on the number of nodes. Hence the runtimes for every face vary drastically, which leads to unbalanced parallelization and low efficiency.

Load balancing algorithms solve this issue. These algorithms use an estimate for the runtimes and partition the tasks into several sub set. Depending on the constraints and fixed or variable number of processors, these partitioning problems are sorted to different problem classes. But all of them belong to the NP (nondeterministic polynomial time) problem class, refer to work by Johnson [13]. Hence solutions are often only nearly optimal and the algorithms are based on heuristics.

The static load balancing uses runtime estimates for every task and assigns the tasks to the processors. These estimates are obtained by running a subset of all simulation parameters and computing an approximation polynomial from their runtimes. Later on, runtimes are extrapolated based on the remaining number of nodes. The aim is a short overall runtime with the least number of processors as possible. The algorithm for a nearly optimal approximate partitioning is similar to the first-fit decreasing algorithm, as mentioned in [14]. In this algorithm every 'bin' represents a processor and faces are added into the bins. All faces in one bin are computed sequentially. Originally the algorithm requires a fixed bin size, which would correspond to the maximal runtime per core. But that size has to be minimized too and is not known a priory. Hence the bin selection has to be changed from "fit-first" to "smallest work".

The algorithm starts with sorted faces with respect to their estimated work in decreasing order. Afterwards the faces are assigned from highest to lowest work sequentially to the bin with lowest work. Whereas the original first-fit algorithm requires a fixed bin size, this strategy tries to keep the bin sizes minimal. The number of processors corresponds to the number of bins, which is to be minimized too. That minimization can be approximated by trying an increasing number of bins and stop, if the total estimated runtime does not decrease.

If the runtime, or the maximum work of the bins, is less than the runtime in the previous iteration with a smaller number of bins, the number of bins is increased and the assignment gets restarted. The implementation of static load balancing will be illustrated in the next section where machine tool faces under investigation will be assigned to bins depending on the runtime of clustering.

## 5. CASE STUDY- REDUCTION OF THE COMPUTATIONAL TIME WITH (DISTRIBUTED) PARALLELIZATION

5.1 Simulation Model

The basic idea or motive behind parallelization of decoupling simulation approach is to reduce the computation time involved in two processes:

   i.   CFD Simulations in ANSYS CFX
   ii.  Optimal subset search in MATLAB.

The extend to which this approach is effective can only be identified by performing the operations on a sufficiently complex and justifiable simulation model. The geometry chosen for this investigation is the machine tool- Auerbach ACW 630, a three-axis milling machine of the Chemnitz University of Technology. The motive behind every scientific investigation is to obtain results in close acceptance to real-life scenarios. The simulation model chosen for this investigation involves the machine tool within an octagonal flow environment. Octagonal faces facilitate a multitude of flow directions around the machine tool. However, for the current case study, only one flow direction will be utilized with only ambient air temperature and air velocities varying.

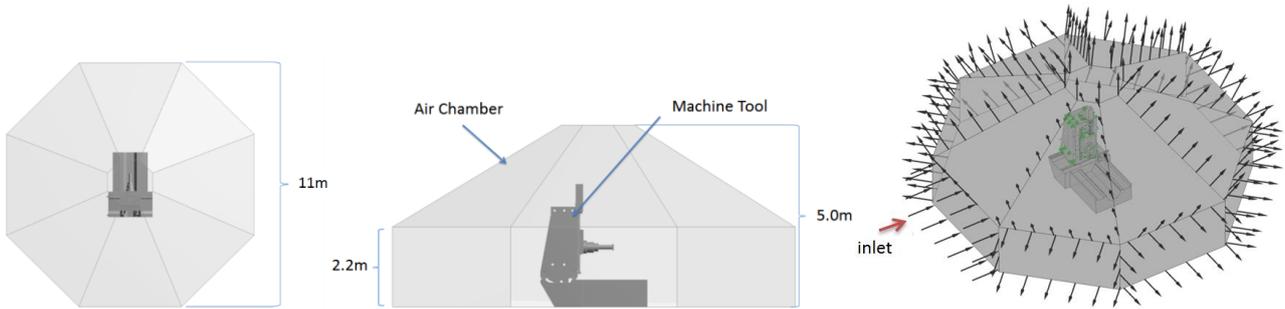

Fig. 4. Octagonal flow chamber with prismatic roof

The parameterized environmental influences included : [Air Temperature (°C), Inlet Velocity (m/s), Azimuth (degree) and Elevation(degree) ]. Azimuth and elevation angles were maintained '0°' throughout the load cases, suggesting the inlet flow to be from the left-most face of the octagonal chamber. The air temperature was varied from 10°C to 40°C with intervals of 5°C and inlet velocity varied from 1 m/s to 9 m/s with intervals of 2 m/s. Each temperature was mapped to all possible combination of velocities, thereby creating 35 (5*7) load cases in total.

The final FE mesh revealed 1483,864 elements and 554,053 nodes. Heat sources are defined at motor positions, friction guides and slides with experimentally recorded values. The inlet is from the left and all the remaining faces act as outlet, as shown in figure 4.

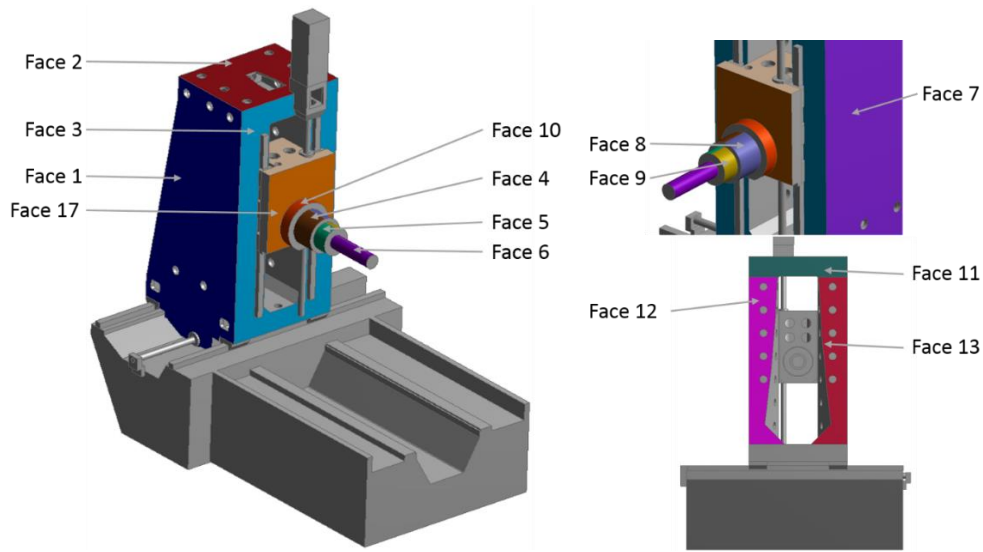

Fig. 6. (Some) selected faces for HTC data export

Parallelization 1 performs simultaneous simulations for each combination of load cases. The attributes related to parallelization 1 is mentioned in table 1. The most prominent outer faces (see figure 6) on the machine column, which are expected to have the most thermal interaction with the surrounding air and most influence on TCP- displacement are chosen for HTC-export after CFD simulations. The HTC values at these selected faces (figure 6) are exported as *.csv files using a workbench journal script file which also involves CFX Command Language and CFX Expression Language.

Table 1. Attributes and procedure: Parallelization 1

| Architecture | Haswell |
|---|---|
| Manufacturer | Intel |
| Memory upper bound | 4069 MB |

5.2  Comparison and quantification of reduction in runtime for simulation decoupling.

The previously described simulations are uniquely defined by the geometry and the surface velocity. Hence each simulation is independent from the others and they can be run simultaneously. From a technical point of view the simulations are controlled by the ANSYS scripting engine, refer AnsysACT [15] and a job queue which manages the resources.

The figure 7 depicts the runtimes of the simulations for every experiment with blue bars and the operational time for copy and wait in red and orange respectively. All jobs start nearly immediately which is more of an exception. The organization and internal setup of ANSYS require a copy of every simulation beforehand. However, the copying times are negligible compared to the runtimes. Therefore, backing up the original model and running every simulation on a backup does not lead to a significant time overhead.

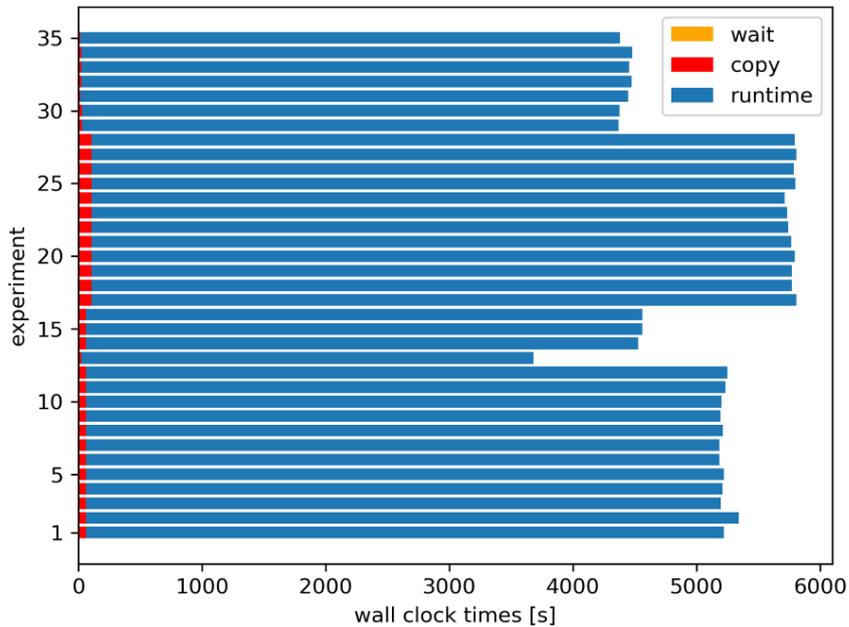

Fig.7. Parallel runtime of the simulation together with the management operations for project file copy and batch system wait times.

On the first sight at figure 7 a very strong variation in the simulation times can be observed between 6000 seconds and 4500 seconds. This can have several reasons. One reason is the heterogenous structure of HPC systems at ZIH/TU-Dresden (refer [16]).

The batch system, which manages the job queue, selects the nodes according to the usage and therefore might select nodes with older CPUs. On the other hand every simulation runs with another parameter set and the underlying numerical methods might be very sensitive to these parameters too. Nevertheless with all simulations run in parallel, hence the total wall clock time is below two hours. In contrast, a complete sequential simulation of all 35 experiments would have lasted more than 177914s ~ 50h. In other words the parallel approach led to a good speedup about 31 with 35 cores.

### 5.3 Runtimes of statically balanced parallelization

Parallelization 2 presents a new set of challenges since the runtimes for each clustering operation per machine face can vary drastically from each other. As stated in the load-balancing algorithm, runtime estimates are required for the assignment. From an efficiency point of view, faces with smallest runtime are selected along with a small number of medium sized faces and their runtimes are measured. From these runtimes, the approximation polynomial as shown in figure 8 can be obtained.

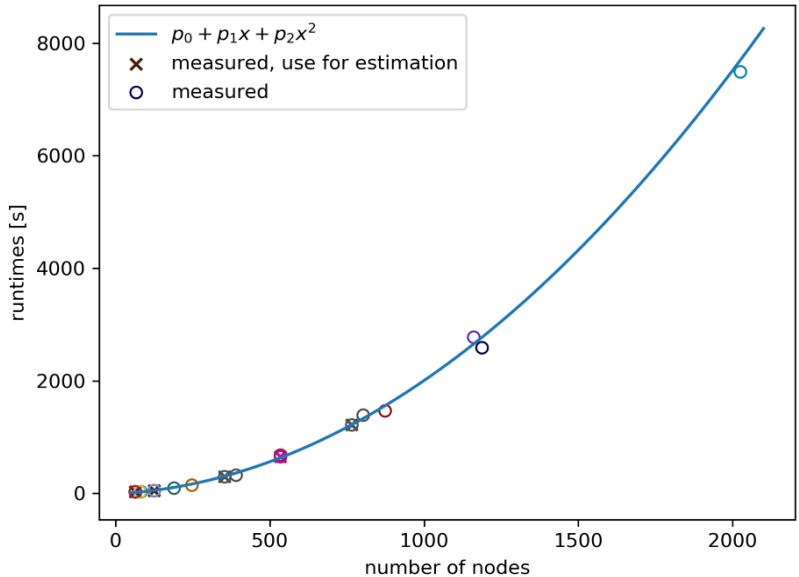

Fig.8. the runtimes for some faces and a quadratic polynomial fit. (cross denotes the faces used for estimation, whereas the circles depict all runtimes.

The measured runtimes for some faces and the quadratic fitting polynomial is given in figure 8. There were five faces used for runtime estimation. The corresponding quadratic polynomial also estimated the runtimes for the other faces with acceptable precision.

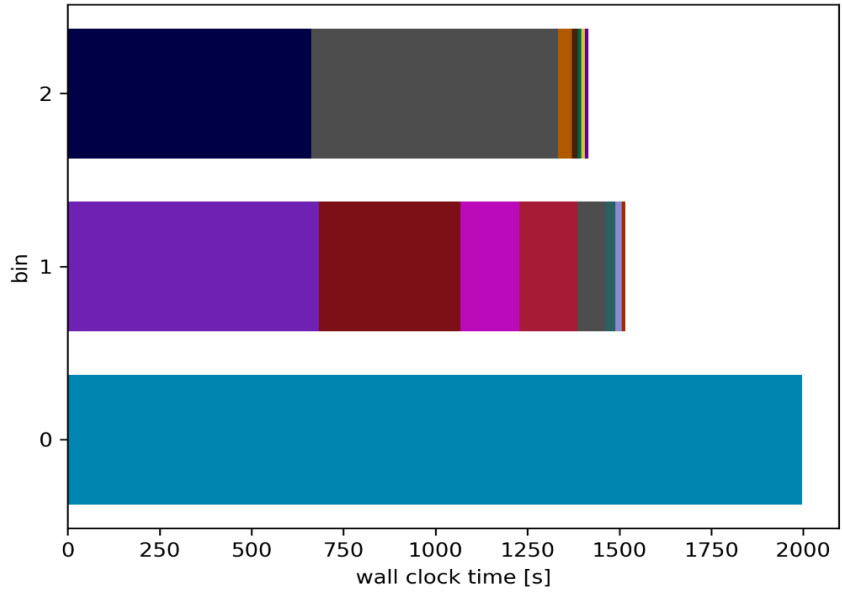

Fig.9. The runtimes for the three bins in Table 2.

As mentioned in section 4, each processor corresponds to a 'bin' and the jobs based on their runtimes are allocated to the bins such that a load balance is attained. The balancing algorithm led to the bins in Table 2. Each color in figure 9 corresponds to a face in figure 6. Hence, the largest face leads to the largest runtime and determines the bin sizes. The face with the most number of FE-nodes fills one bin completely and gets computed in parallel to all other faces, whereas the smaller faces were assigned to the last two bins. As can be seen in Figure 9 the runtimes are all nearly balanced. Running all experiments sequentially would require a runtime about 5082 seconds, whereas the parallel method required 2043 seconds. This amounts to a speed up factor of 2.48 with three cores.

Table 2. The bins and their assigned faces.

| Bin number | Faces |
|---|---|
| 0 | 3 |
| 1 | 1, 15, 14, 18, 17, 4, 5, 9, 6, |
| 2 | 7, 2, 12, 13, 16, 11, 8, 10 |

Thereby, HTC data exported from a limited number of CFD simulations serve as the training data which is further reduced using clustering algorithm and confined to optimal node points obtained on each face of the machine tool. Parallelization of optimal subset search facilitates simultaneous optimal subset search operations and thereby drastically reducing the computation time. The CDs are trained based on these optimal subsets obtained through parallelization 2. As mentioned before, convection data is interpolated for any combination of environmental load cases based on the training data using CDs. This serves as the boundary data for thermo-elastic simulations. This approach increases in prominence, when moving machine components or changes in flow direction are considered. Clustering and training data generation in such scenarios would be extremely time-consuming if parallelization is not utilized.

## 6. SUMMARY, CONCLUSION AND OUTLOOK

Decoupling of fluid simulations from thermo-mechanical simulations is of growing significance in thermal-error prediction because of the complexities of the geometries and dynamic behaviours of machine parts involved. For each variation in environmental conditions around a machine tool, it is impractical to perform highly time-consuming CFD simulations for boundary data in order to ascertain the corresponding thermal displacements. Decoupling approach predicts convection data for thermo-mechanical simulations based on trained characteristic diagrams. Clustering algorithm accelerates the training operation by efficiently optimizing the training data. Holistically however, the decoupling approach needs to be quickened up by multiple factors in order to be incorporated as a reliable thermal-error prediction method.

This paper attempts to overcome the shortcomings of the decoupling approach in terms of computation time. Parallel computation is incorporated in two stages, which leads to the training of CDs. Firstly, CFD simulations in ANSYS CFX are run in parallel utilizing the independence of load cases among each other and almost identical runtimes for each simulation. On recent high performance clusters, every simulation can run in parallel, leading to nearly optimal speed ups. With 35 cores a good speed up of about a factor of 31 was obtained.

The second stage of parallelization, involves optimal subset search on each face of the machine tool. The fitness function of GA represents the expensive function, which is solved in parallel. Due to extremely different runtimes for each face, because of varying number of FE-nodes 'static load balancing' algorithm is incorporated to efficiently place the numerical experimental or jobs in 'bins' or processors. This optimal load balancing yields a speed up of about 2.5 with three cores.

Both parallelization operations can be improved by an additional parallelization level. The CFD simulations can fully employ the parallel advantages of distributed finite volume methods. Using parallel linear algebra structures, the evaluation of fitness function can be quickened up further. Both approaches require a sophisticated experimental organization on the high performance clusters.

The future scope of work would include parallelization in decoupling for a dynamic machine, where positions of the moving components should also be considered as input parameters in CDs. The environmental influences could also be defined as functions of time if the movement of machine components are pre-known. All possible flow directions should be incorporated as well and evaluation times compared.

ACKNOWLEDGEMENTS

*This research was supported by a German Research Foundation (DFG) grant within the Collaborative Research Centers/Transregio 96, which is gratefully acknowledged.*